\documentclass[letter]{aa}
\usepackage{graphicx}
\usepackage{txfonts}
\usepackage{float}
\usepackage{subfig}
\usepackage{amsmath}
\usepackage{comment}

\def\kms{km\,s$^{-1}$}
\def\zgt6{\hbox{{\it z}$\,>\,$6}}

\def\lsim{\mathrel{\rlap{\lower 3pt \hbox{$\sim$}} \raise 2.0pt \hbox{$<$}}}
\def\gsim{\mathrel{\rlap{\lower 3pt \hbox{$\sim$}} \raise 2.0pt \hbox{$>$}}}

\begin{document}

\authorrunning{Castignani et al.}
\titlerunning{Molecular gas observations of the recoiling black hole candidate 3C~186}

\title{{NOEMA observations support a recoiling black hole in 3C~186}}

\author{G. Castignani
          \inst{1,2}\fnmsep\thanks{e-mail: gianluca.castignani@unibo.it}
          \and
        E. Meyer\inst{3} 
        \and          
         M. Chiaberge\inst{4,5}
         \and
         F. Combes\inst{6,7}
        \and
        T. Morishita\inst{4,8}
        \and
        R. Decarli\inst{2}
        \and
        A. Capetti\inst{9}
        \and
        M. Dotti\inst{10,11}
        \and
        G.~R. Tremblay\inst{12}
        \and
        C.~A. Norman\inst{5,13}}

   \institute{Dipartimento di Fisica e Astronomia ''Augusto Righi'', Alma Mater Studiorum Università di Bologna, Via Gobetti 93/2, I-40129 Bologna, Italy   
            \and
        INAF - Osservatorio  di  Astrofisica  e  Scienza  dello  Spazio  di  Bologna,  via  Gobetti  93/3,  I-40129,  Bologna,  Italy
            \and
        University of Maryland Baltimore County, 1000 Hilltop Circle, Baltimore, MD 21250, USA
        \and
        Space Telescope Science Institute for the European Space Agency (ESA), ESA Office, 3700 San Martin Drive, Baltimore, MD 21218, USA
        \and
    The William H. Miller III Department of Physics \& Astronomy, Johns Hopkins University, Baltimore, MD 21218, USA
        \and
         Observatoire de Paris, PSL University, Sorbonne University, CNRS, LERMA, F-75014, Paris, France
         \and
         Coll\`{e}ge de France, 11 Place Marcelin Berthelot, 75231 Paris, France          
         \and
        Infrared Processing and Analysis Center, California Institute of Technology, MC 314-6, 1200 E. California Boulevard, Pasadena, CA 91125, USA
         \and
         INAF - Osservatorio Astrofisico di Torino, Via Osservatorio 20, I-10025 Pino Torinese, Italy
         \and 
         Universit\`{a} degli Studi di Milano-Bicocca, Piazza della Scienza 3, 30126 Milano, Italy
         \and
         INFN, Sezione di Milano-Bicocca, Piazza della Scienza 3, 30126, Milano, Italy
         \and
         Harvard-Smithsonian Center for Astrophysics, 60 Garden Street, Cambridge, MA 02138, USA
         \and
         Space Telescope Science Institute, 3700 San Martin Dr., Baltimore, MD 21210, USA
                 }            
             
\date{Received: February 11, 2022; Accepted: April 12, 2022}

\abstract{3C~186 is a powerful radio loud quasar (QSO) at the center of a cool-core cluster at $z=1.06$. Previous studies reported evidence for a projected spatial offset of $\sim1^{\prime\prime}$ between the isophotal center of the galaxy and the {point-source} QSO {as well as} a spectral shift of $\sim2000$~km~s$^{-1}$ between the narrow  and broad line region of the system. In this work we report high-resolution molecular gas CO(4$\rightarrow$3)  observations of the system taken with NOEMA interferometer. We clearly detect a large reservoir of molecular gas, $M_{H_2}\sim8\times10^{10}~M_\odot$, that is cospatial with the host galaxy and likely associated with a {rotating disk-like structure.} {We firmly confirm both the spatial offset {of  the galaxy's gas reservoir} with respect to the continuum emission of the QSO and the spectral offset with respect to the redshift of the broad line region.} Our morphological and kinematical analysis  {confirms that the most likely scenario to explain the 3C186 system is that the QSO is a kicked super-massive black hole} (SMBH), which we believe may have resulted from a strong gravitational wave recoil as two SMBHs coalesced after the merger of their host galaxies.}






\maketitle

\section{Introduction}
Galaxy mergers are fundamental mechanisms regulating galaxy growth and the evolution of super massive black holes (SMBH) at their centers. SMBH mergers are expected to be among the most energetic phenomena in the Universe, where a small fraction of the SMBH binding energy $\frac{\Delta E}{M_{\bullet} c^2} \simeq 10^{-5}(\frac{M_{\bullet}}{10^6 M_\odot})^{1/4}$  is converted into a large amount $\sim10^{55}~{\rm erg}~(\frac{M_{\bullet}}{10^6 M_\odot})^{5/4}$ of gravitational wave (GW) radiation \citep{Colpi2014}, where $M_{\bullet}$ is the black hole mass. Depending on both the  relative orientation of the spins of the merging SMBHs and their mass ratio, the merged SMBH may receive a recoil kick \citep[][for some reviews]{Centrella2010,Komossa2012} with velocities as high as $\sim$5000~km~s$^{-1}$
\citep{Campanelli2007,Lousto_Zlochower2011}, resulting from highly
anisotropic GW radiation. With such a high velocity, the SMBH may be {measurably} displaced from the center of the host galaxy.


It is still debated how two SMBHs could reach the distance at which GW losses become important, the so-called final-parsec problem \citep[e.g.,][]{Milosavljevic_Merritt2003}. It
is possible that SMBH pairs may stall and never merge, although a supply of torques from ambient gas may help to overcome the problem. Simulations also show that even in gas-poor environments SMBH
binaries can merge under certain conditions, e.g. if they formed in major galaxy mergers where the final
galaxy is non spherical \citep{Bortolas2016, Khan2012,Preto2011}. Observational
evidence of unambiguous cases of galaxies hosting GW recoiling SMBHs would have tremendous impact on our understanding of this process. Confirmed candidates are also important for future space-based missions such as eLISA and pulsar-timing experiments \citep{AmaroSeoane2012,Klein2016,Kelley2018}. They are sensitive to GWs of lower frequency (down to nHz) than those associated with stellar-mass black hole mergers (from Hz to kHz) detectable from ground-based interferometers such as Virgo and LIGO \citep{Abbott2021a,Abbott2021b}. A confirmed recoiling SMBH candidate would provide evidence that SMBHs do in fact merge.

To that end, this work focuses on {the} recoiling SMBH candidate {recently} discovered in the radio-loud
quasar 3C~186 \citep[][]{Chiaberge2017}, at the center of a distant cool-core X-ray cluster at $z=1.06$ \citep{Siemiginowska2005,Siemiginowska2010}.
{\it Hubble Space Telescope (HST)}  imaging shows that the bright quasar (QSO) in this system is offset by $1.3^{\prime\prime}\pm0.1^{\prime\prime}$ with respect to the host galaxy isophotal center \citep{Hilbert2016}. New deep {\it HST}/WFC3-IR and ACS images confirm this spatial offset at high significance \citep{Morishita2022}. Furthermore, integral field spectroscopy with Keck shows evidence for blueshifted broad emission lines, with a significant velocity offset of {$\sim2000$~km~s$^{-1}$} with respect to the narrow emission lines \citep{Chiaberge2018}. 

It seems most likely that the narrow lines trace the more quiescent kinematics of the host galaxy. {These lines are indeed produced in the narrow line region (NLR) on kiloparsec scales, far from the SMBH. Conversely,} the broad line region (BLR) remains coupled to the recoiling SMBH \citep{Loeb_Wyithe2008}, and thus the high relative velocity can be linked to the offset SMBH. 
{An interpretation that self-consistently explains} both spatial and spectral offsets is that the SMBH in 3C~186 is recoiling as a result of a gravitational radiation rocket effect following a major galaxy merger in which the central SMBHs merged. High-velocity ($>$1000~km~s$^{-1}$) kicks are expected to be rare, but they are {also} more likely to be observed in combination with large spatial offsets such as the one we observe in 3C~186 \citep{Lousto2012,Blecha2016}.

If the GW recoil following a merger of two SMBHs is the correct interpretation, \citet{Lousto2017} showed that this would be the most energetic event ever observed: a major merger of progenitor SMBHs of comparable mass, giving rise to a highly spinning SMBH ($a=0.91$), where $\sim9\%$ of the total mass is radiated as GW.
According to the GW-recoiling black hole scenario for 3C~186, we expect most of the {molecular gas, which is the fuel of star formation,} to have a redshift consistent with the narrow lines that we believe rest in the host galaxy frame {\citep{Chiaberge2017}.}


\citet{Chiaberge2017} already tested some alternatives to the GW recoil scenario: a dual active galactic nucleus (AGN), a pair of two SMBHs, a double-peaked disk \citep[e.g.,][]{Eracleous_Halpern2003}, or a quasar with extremely fast winds. However, {none adequately}
explain both the spatial and the spectral offsets.
Furthermore, recent deep {\it HST} observations by \citet{Morishita2022}, supported by simulations, {appear} to exclude the possibility of a recent/ongoing merger.

The main goal of this work is to test  the {GW recoil} hypothesis by mapping  the cold gas reservoir in the 3C~186 system and determining its redshift. Similar observations of those presented in this work  were vital to conclusively refute the recoiling SMBH scenario in a different candidate source  \citep[J0927+2943 at $z = 0.7$,][]{Decarli2014}.
The present study is part of a larger campaign to study the environmental processing of distant (brightest) cluster galaxies via CO observations \citep{Castignani2018,Castignani2019,Castignani2020a,Castignani2020b,Castignani2020c,Castignani2020d}.


Throughout this work we adopt a flat $\Lambda \rm CDM$ cosmology with matter density $\Omega_{\rm m} = 0.30$, dark energy density $\Omega_{\Lambda} = 0.70$ and Hubble constant $h=H_0/100\, \rm km\,s^{-1}\,Mpc^{-1} = 0.70$. The luminosity distance at the redshift $z=1.06$ of 3C~186 is  7,101~Mpc { (i.e., $1'' = 8.11$~kpc).} 



\section{NOEMA 1.8~mm observations and data reduction}\label{sec:obs}

We observed 3C~186 with NOEMA  on {1, 5, and 10} March 2021, in A configuration, as part of the program W20CO (PI:~Castignani). We used 10 antennas on 5 March, and 11 antennas during the other two days of observations. With 10 and 11 antennas configurations the time spent on-source was 3.4~hr and 8.2~hr, respectively. The total observing time was 16.2~hr (i.e., 11.6~hr on source).

We set the phase center of the observations equal to 3C~186 QSO coordinates R.A.~=~7$^{\rm h}$:44$^{\rm m}$:17.5$^{\rm s}$ and Dec.~=~37$^\circ$:53$^\prime$:17.2$^{\prime\prime}$. 
We used the PolyFix correlator which covers a total bandwidth of 15.5~GHz, in each linear polarization, split between the lower and upper side bands.
We adopted a tuning frequency  in the upper inner (UI) baseband of 223.7~GHz, which is the mean between the redshifted CO(4$\rightarrow$3) observed frequencies $\nu_{\rm obs}$ for the NLR ($z=1.068$, $\nu_{\rm obs}=222.940$~GHz) and the BLR ($z=1.054$, $\nu_{\rm obs}=224.459$~GHz). With this tuning the CO(4$\rightarrow$3) lines of both BLR and NLR thus fall within the 3.9~GHz wide UI baseband. 

The baseline range was 32--760\,m. The program was executed in average winter weather conditions, although unstable, with a 
system temperature $T_{\rm sys}=120-250$\,K  and a precipitable water vapor column of $\sim1-3$\,mm. The sources 0805+410 and 0738+313 were used as phase and amplitude calibrators,  while 3C~84 and 3C~279 served as bandwidth and flux calibrators. We reduced the data using \textsf{clic} package of the \textsf{GILDAS} software, to obtain the final ($u$,$v$) tables.

We then imaged the visibilities using the \textsf{MAPPING} software  of \textsf{GILDAS}. At the tuning frequency of 223.7~GHz, the half power primary beam width is $22.5''$. We adopted natural weighting, yielding a synthesized beam of $0.62''\times0.42''$ with PA=14$^\circ$. We then re-binned the spectral axis at a resolution of 53.61\,\kms{} (i.e., 40\,MHz at 223.7 GHz). {At this resolution the resulting {root mean square} (rms) is 0.35\,mJy\,beam$^{-1}$.} 
As mentioned above, the frequency tuning was set to 223.7~GHz for our observations. However, in the following Sections we will always report velocities relative to the redshift of the NLR. With this convention the CO(4$\rightarrow$3) line for the NLR and BLR correspond to systemic velocities equal to 0~km~s$^{-1}$ and -2035.7~~km~s$^{-1}$, respectively.


\begin{table*}[htb]
\begin{center}
\begin{tabular}{lccccccc}
\hline
Component & $S_{\rm CO(4\rightarrow3)}$ & FWHM & $L^\prime_{\rm CO(4\rightarrow3)}$ &  $z$ & velocity \\
 & (Jy~km~s$^{-1}$) & (km~s$^{-1}$) & ($10^{9}$~K~km~s$^{-1}$~pc$^2$) &  & (km~s$^{-1}$) \\
\hline
Total & $2.09\pm0.11$  & $269\pm17$ & 7.94$\pm$0.42 & $1.06811\pm0.00005$ & $16\pm7$ \\ 
North & {$0.20\pm0.03$} & {$267\pm47$}  & {$0.76\pm0.11$}  & {$1.0684\pm0.0004$} & {$57\pm22$}  \\ 
South & {$0.22\pm0.03$} & {$304\pm52$}   & {$0.84\pm0.11$}  & {$1.0685\pm0.0001$} & {$76\pm19$} \\




\hline
 \end{tabular}
\end{center}
 \caption{CO(4$\rightarrow$3) results for the total emitting region, denoted as {\it Total}, as well as for the northern {and southern components.}} 
\label{tab:COresults}
\end{table*}




\begin{figure}
\begin{center}
 \includegraphics[width=0.44\textwidth]{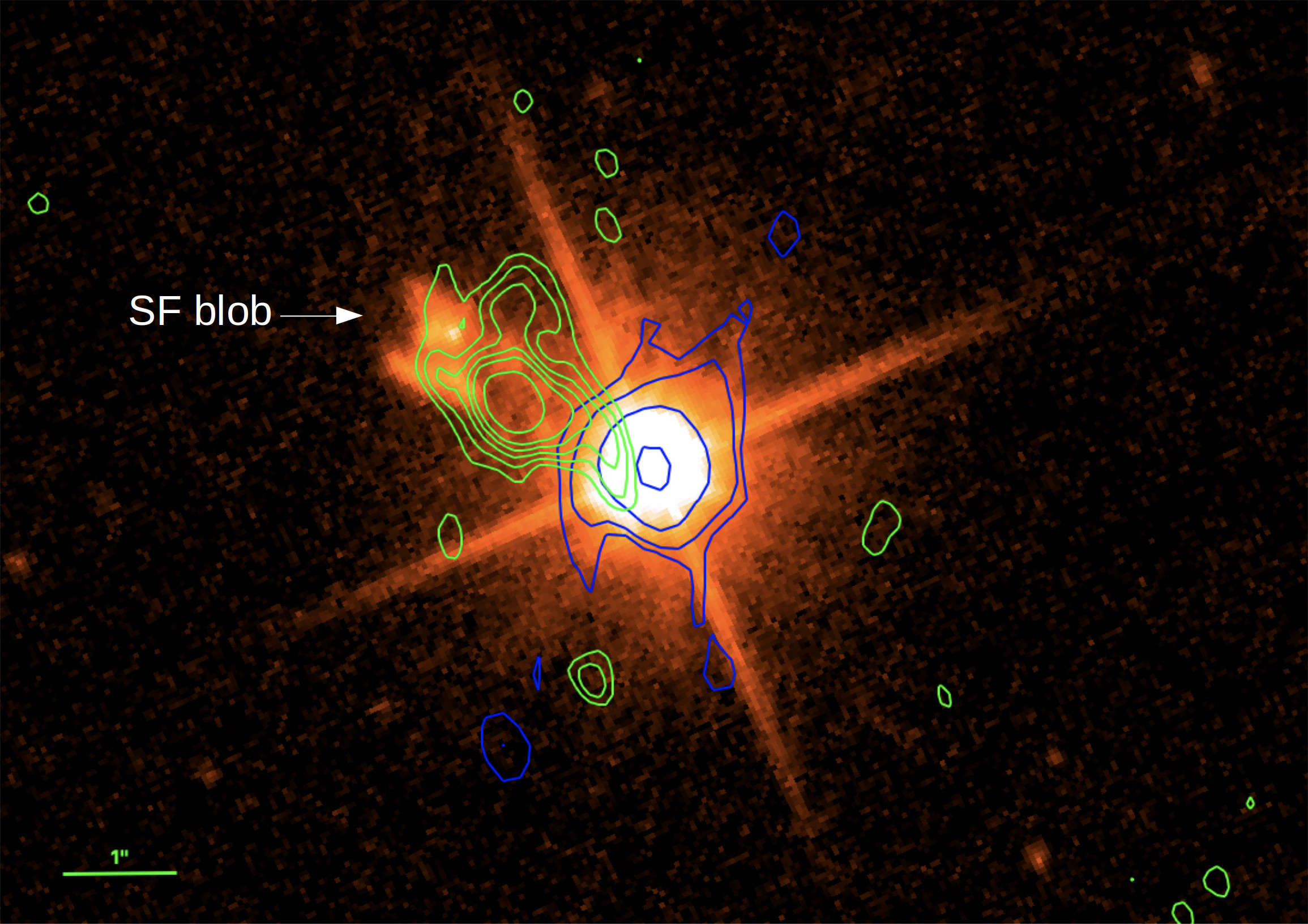}
\end{center}
 \caption{{\it HST} WFC/F606W image (rest-frame UV) of 3C~186. Blue contours show the 1.3~mm continuum emission from the {QSO}, at levels of 0.10, 0.16, 0.40, and 1.34~mJy~beam$^{-1}$. Green contours correspond to the CO(4$\rightarrow$3) emission at levels of 
{183, 244, 305, 366, 427, and 549}~mJy~km~s$^{-1}$~beam$^{-1}$. The CO(4$\rightarrow$3) emission is at the NLR redshift, redshifted by $\sim2000$~km~s$^{-1}$ with respect to the BLR that is cospatial with the quasar. North is up, East is left. The SF blob at NE is highlighted.}
\label{fig:HST_image}
\end{figure}

\section{Results}\label{sec:results}
\subsection{CO(4$\rightarrow$3) emission}

We clearly detect extended $2^{\prime\prime}\times2^{\prime\prime}$ CO(4$\rightarrow$3) emission along the north-east direction at the NLR redshift with {a peak} reached $0.6^{\prime\prime}$ to the north and $1.3^{\prime\prime}$ to the east from the quasar, at the coordinates (R.A.; Dec.) = ($7^{\rm h}44^{\rm m}17.58^{\rm s}; 37^\circ53^\prime17.8^{\prime\prime}$).

{Figure~\ref{fig:HST_image} displays the {\it HST} WFC/F606W image of 3C~186, in its rest-frame ultraviolet (UV). The central QSO is cospatial with the continuum emission at 1.3~mm (blue contours).  The extended CO(4$\rightarrow3$) emission is shown as green contours. Its peak is co-spatial with the isophotal center of the host galaxy \citep[determined from 2D galaxy modeling,][]{Chiaberge2017}, and clearly offset by $1.4^{\prime\prime}$ to the NE from the central QSO. The CO emission extends down to the QSO itself, in projection. It also partially overlaps with a star forming (SF) blob at the NE \citep{Hilbert2016}, highlighted in Fig.~\ref{fig:HST_image}.}

Figure~\ref{fig:maps}, panels (a, b, c), shows  moment 0, 1, and 2 maps  that display the CO(4$\rightarrow$3) detection and are obtained after subtracting the continuum emission. The intensity map shows the CO(4$\rightarrow$3) emission, with its peak clearly shifted by $1.4^{\prime\prime}$ to the north-east from the 3C~186 quasar coordinates. The total emitting region is partially resolved by our observations. The intensity map shows evidence for a northern component  connected to the main emitting region and a southern trail (hereafter denoted as southern component) which extends to the south-west almost reaching the {position of the} 3C~186 quasar. These two components are denoted as {\it N} and {\it S} in Fig.~\ref{fig:maps}a. {They do not only show up as morphological extensions of the main CO emitting region in the intensity map, but they also appear to have distinct kinematics, as further discussed in Sect.~\ref{sec:CO_kinematics}. In the velocity map (Fig~\ref{fig:maps}b) we highlight both {\it N} and {\it S} components, centered at (R.A.; Dec.) = (7$^{\rm h}$44$^{\rm m}$17.56$^{\rm s}$; 37$^\circ$53$^\prime$18.7$^{\prime\prime}$) and (7$^{\rm h}$44$^{\rm m}$17.52$^{\rm s}$;37$^\circ$53$^\prime$17.4$^{\prime\prime}$), respectively.}

In Fig.~\ref{fig:CO43spectrum} (panel a) we report the CO(4$\rightarrow$3) spectrum for the total emitting region. In panels (b) and (c) we report CO(4$\rightarrow$3) spectra for the north and south components, separately. For the total system our Gaussian fit yields a high signal-to-noise (S/N) ratio of 19, corresponding to a velocity integrated flux ${S_{\rm CO(4\rightarrow3)}\Delta v=(2.09\pm0.11)}$~Jy~km~s$^{-1}$, a full width at half maximum ${{\rm FWHM}=(269\pm17)}$~km~s$^{-1}$, and a redshift of $1.06811\pm0.00005$. The redshift is fully consistent with that of the narrow forbidden emission lines of the host galaxy, from the literature \citep[i.e., $z= 1.0685\pm0.0004$,][]{Chiaberge2017,Hewett_Wild2010}. 

Our observations are sensitive to the CO(4$\rightarrow$3) line from sources within 22.5~arcsec radius (i.e., 180~kpc) from the 3C~186 QSO, {while the upper side band covers a broad velocity range $\sim(-8,000;+2000)$~km~s$^{-1}$ with respect to the host galaxy (i.e., the NLR).} {Therefore, additional lines could have been serendipitously detected from companion cluster members.} However, by inspecting the entire datacube, no other emission line was detected in the upper side band, neither in the lower side band. The two bands are sensitive to CO(4$\rightarrow$3) from sources at redshifts $z=1.01-1.08$ and $z = 1.16-1.24$, respectively.


\subsection{CO(4$\rightarrow$3) kinematics}\label{sec:CO_kinematics}
The velocity map reported in Fig.~\ref{fig:maps}b shows that the {bulk} of the emitting region has a negligible velocity shift with respect to the NLR of  the host galaxy. {Both north and south components  have instead a velocity up to $\sim$+100~km~s$^{-1}$ relative to the NLR, which implies that these two slightly recede along the line of sight with respect to the main emitting region, while being mainly distributed in the projected space.} Such a moderate velocity shift for the two components implies a velocity gradient along the east-west direction which suggests the presence of a mildly inclined {disk-like structure,} which however shows complex and disturbed morphology and kinematics. 



The moment 2 map in Fig.~\ref{fig:maps}c shows that there are no strong gradients in the velocity width over the projected extension of the emitting region. {The CO(4$\rightarrow$3) line width ranges between $(189^{+309}_{- 126})$~km~s$^{-1}$ over the entire emitting region, where the median value is reported, along with the 68$\%$ interval. The highest values are found close to the location of the intensity peak, as in Fig~\ref{fig:maps}a. This result is consistent with the presence of a star forming disk-like structure.} Indeed, the molecular gas reservoir typically shows an exponentially declining density profile, thus quite concentrated at the galaxy center \citep{Nishiyama2001,Regan2001,Leroy2008}.

\subsection{Continuum emission}\label{sec:continuum}
{By using the entire lower and upper side bands (LSB, USB), separately, we looked for continuum emission, which we clearly detect in both sidebands, with similar flux densities. 
Figure~\ref{fig:maps}d and the blue contours of Fig.~\ref{fig:HST_image} display the continuum map in the LSB, which reveals a strong point-like component down to a 5$\sigma$ significance level (rms = 0.03~mJy~beam$^{-1}$), co-spatial with the 3C~186 QSO. There is also evidence of more extended emission down to a 3$\sigma$, which appears to be elongated along the NW-SE direction, similarly to the larger scale VLA jet \citep{Hilbert2016}. By integrating over the region around the QSO delimited by the 5$\sigma$ and 3$\sigma$ contours we find flux densities of $2.7\pm0.1$~mJy and $3.7\pm0.2$~mJy, respectively. This implies that the point-like source is dominant and accounts for $\sim70\%$ of the total emission around the QSO.} 

{Fig.~\ref{fig:maps}e displays the cleaned map of the residuals, where we removed the central point-like component in the  $uv$-visibility plane, assuming a Gaussian model.  A fit with \textsf{MAPPING} yields  a full width at half power ${\rm FWHP}=0.28^{\prime\prime}$, thus similar to beam size. Extended features at a level of 3$\sigma$ around the QSO are visible in the map, likely originated by the jet emission. The non-thermal origin of the observed continuum emission is further strengthened when we compare the 1.3~mm continuum flux density with the total flux density of $(1.31\pm0.07)$~Jy at 1.4~GHz \citep{Laing_Peacock1980}. We infer a spectral index $\alpha=1.2\pm0.1$, typical of  synchrotron emission $S_\nu\propto\nu^{-\alpha}$ of steep spectrum radio loud quasars as 3C~186.} 

Furthermore, the QSO is cospatial with a flat-spectrum component (K1), identified 
by \citet{Spencer1991} via high-resolution $\lesssim0.03^{\prime\prime}$ observations at 1.6~GHz, with a corresponding flux density of $\sim$12~mJy.
Our observations have instead a spatial resolution of $\sim0.5^{\prime\prime}$ so that we cannot spatially resolve the non-thermal emission of the core from that of the one-sided jet along the north-west direction \citep{Spencer1991}.  However, combining the  1.6~GHz (K1) and 1.3~mm continuum emissions {still yields a steep spectrum with $\alpha\simeq0.3$, although  flatter than that obtained using the total 1.4~GHz flux density. All these results suggest} that the 1.3~mm continuum emission likely includes the contribution of both the core and the more extended jet.



\subsection{Molecular gas reservoir}\label{sec:result_MH2}

{We estimate} a total velocity integrated luminosity of ${L^\prime_{\rm CO(4\rightarrow3)}=(7.94\pm0.42)\times10^{9}}$~K~km~s$^{-1}$~pc$^2$, using the following formula taken from \citet{Solomon_VandenBout2005}:
\begin{equation}
{\small \label{eq:LpCO}
\frac{L^{\prime}_{\rm CO(4\rightarrow3)}}{{\rm K}~{\rm km}~{\rm s}^{-1}~{\rm pc}^2}=1.53\times10^2\,\frac{S_{\rm CO(4\rightarrow 3)}\,\Delta\varv}{{\rm Jy}~{\rm km}~{\rm s}^{-1}}\,\bigg(\frac{D_L}{\rm Mpc}\bigg)^2\,(1+z)^{-1}\;,}
\end{equation}
where $D_L$ is the luminosity distance at the redshift $z$ of the source. Similarly, we have derived molecular gas properties for the north and south components, separately, from the fits reported in Fig.~\ref{fig:CO43spectrum}.

Each of these two components is clearly detected at {$\mathrm{S/N}\simeq7$ and accounts for $\sim10\%$} of the total observed emission.
The best fit results are summarized in Table~\ref{tab:COresults}. As reported in the Table,   the total, north, and south components have similar redshifts, {fairly} consistent with each others within the errors bars, and similar line widths in the range {$\sim269-304$~km~s$^{-1}$} typical of massive galaxies. However, 
as further discussed in Sect.~\ref{sec:CO_kinematics}, {both north and south components have higher recession velocities than the total emitting system.}

{We then estimate} the $H_2$ molecular gas mass as 
 $M_{H_2}=\alpha_{\rm CO}L^{\prime}_{{\rm CO}(4\rightarrow3)}/r_{41}$, by assuming an excitation ratio of $r_{41} = L^{\prime}_{\rm CO(4\rightarrow3)}/L^{\prime}_{\rm CO(1\rightarrow0)}=0.41$ \citep{Bothwell2013,Carilli_Walter2013}, typical of sub-mm galaxies. We adopt a Galactic CO-to-$H_2$ conversion factor $\alpha_{\rm CO}=4.36~M_\odot~({\rm K}~{\rm km}~{\rm s}^{-1}~{\rm pc}^2)^{-1}$, typical of main sequence (MS) star forming galaxies. {This choice is reasonable as the 3C~186 system has a star formation rate (SFR) that is typical of MS galaxies, as further discussed in Sect.~\ref{sec:discussion_conclusions}.}

We obtain a total high $H_2$ gas mass of $(8.4\pm0.4)~10^{10}~M_\odot$ as well as {$M_{H_2}\simeq(8-9)\times10^{9}~M_\odot$  for each of the $N$ and $S$} components. In Fig.~\ref{fig:CO43spectrum}d we report the CO(4$\rightarrow$3) spectrum extracted at the location of the QSO (i.e., the phase center of the observations), showing no emission {line} at the BLR redshift. We used this spectrum to estimate, a resolution of 300~km~s$^{-1}$, a 3$\sigma$ upper limit of $S_{\rm CO(4\rightarrow3)}<0.17$~Jy~km~s$^{-1}$, which corresponds to $M_{H_2}<6.8\times10^{9}~M_\odot$.




\subsection{Dust mass}\label{sec:Mdust}
{\citet{Podigachoski2015} reported evidence for near- to far-infrared emission in 3C~186, between 3.6~$\mu$m and 70~$\mu$m in the observer frame, as well as {\it Herschel} upper limits at longer wavelengths. 
By modeling the spectral energy distribution (SED) the authors then derived a conservative upper limit to the infrared (IR) luminosity of 3C~186,} $L_{\rm IR}<5\times10^{11}~L_\odot$, that we use to set an upper limit to the dust mass \citep[e.g.,][]{Beelen2006}:
\begin{equation}
\label{eq:Mdust}
 M_{\rm dust} = \frac{L_{\rm IR}}{4\pi\int\kappa(\nu)B(\nu,T_{\rm dust})~d\nu}<{1.9\times10^8~M_\odot\;},
\end{equation}
where $\kappa(\nu)=\kappa_0\cdot(\nu/\nu_0)^\beta$ is the dust opacity per unit mass of dust, and $B(\nu,T)$ is the spectral radiance of a black body of temperature T at
frequency $\nu$.

{To derive the $M_{\rm dust}$ upper limit reported in Eq.~\ref{eq:Mdust}} we have adopted $\kappa_0=0.4$~cm$^2$~g$^{-1}$  at $\nu_0 = 250$~GHz \citep[][and references therein]{Beelen2006,Alton2004}, $\beta=1.8$ \citep{Scoville2017}, and a dust temperature $T_{\rm dust}=30$~K, which we inferred using the $T_{\rm dust}$ vs. $L_{\rm IR}$ scaling relation for MS galaxies by \citet{Magnelli2020}.



Under the assumption that molecular gas is shielded by a dust reservoir that is extended over a similar projected area to the CO, a dust continuum emission offset from the QSO would have been {clearly} detectable with our observations, in the Rayleigh-Jeans regime. {As further discussed in Sect.~\ref{sec:continuum} the observed continuum is instead co-spatial with the QSO, with only some $\sim3\sigma$  features towards the CO emitting region.
Integrating the continuum map of the residuals (Fig.~\ref{fig:maps}e) over the extent of the CO emitting region yields a continuum density flux of $\sim0.4$~mJy, which we use to estimate an upper limit to the dust mass $M_{\rm dust}<2.0\times10^8~M_\odot$. To estimate this upper limit we follow  Eq.~A.6 of \citet{Castignani2020a} and adopt the same dust temperature and opacity parameters used above in Eq.~\ref{eq:Mdust}.}

This {continuum-based} $M_{\rm dust}$  upper limit is in excellent agreement with that of Eq.~\ref{eq:Mdust}. We then obtain an $H_2$-to-dust mass ratio of $M_{H_2}/M_{\rm dust}{\gtrsim400}$ for the system, higher than the values $\sim100$, typically found for distant star forming galaxies \citep{Berta2016,Scoville2014,Scoville2016}.

{We stress that AGN emission may affect our estimates leading to biased-high, and thus conservative, $M_{\rm dust}$ upper limits. Lower values would increase the $M_{H_2}/M_{\rm dust}$ ratio and so the tension mentioned above. A lower CO-to-$H_2$ conversion factor than the Galactic one used in this work would instead alleviate the tension. Indeed, assuming an $\alpha_{\rm CO}=0.8~M_\odot~({\rm K}~{\rm km}~{\rm s}^{-1}~{\rm pc}^2)^{-1}$, more typical of starbursts, the $M_{H_2}/M_{\rm dust}$ ratio is reduced by a factor of $\sim5.5$.}

{Alternatively,} to explain the large $M_{{\rm H_2}}/M_{\rm dust}$ ratio it could be that a substantial fraction of the observed gas reservoir has been accreted {by the galaxy}, as a result of a past merger. {Ratios of several hundreds have been in fact commonly found in merger remnants \citep{Davis2015}.} Indeed, \citet{Chiaberge2017} suggested that the 3C~186 system is in a late ($\sim1$~Gyr old) merger phase, from comparison with numerical simulations. {An interpretation that is in agreement with the stellar population age $\gtrsim200$~Myr, as inferred by \citet{Morishita2022} on the basis of a stellar population analysis.} The stellar, gas, and dust properties of the system are listed in Table~\ref{tab:gas_properties}.

\begin{table}[htb]
\begin{center}
\begin{tabular}{l|lcccccc}
$M_\star$ &  $(2.04^{+0.59}_{-0.18})\times10^{11}~M_\odot$  \\
$M_\star$ (NE SF blob) &  $7.94\times10^9~M_\odot$   \\
& \\
SFR$_{\rm IR}$ &  $<80~M_\odot$/yr   \\
SFR$_{\rm SED}$ & $(65\pm20)~M_\odot$/yr   \\
& \\
$M_{H_2}$ &  $(8.4\pm0.4)\times10^{10}~M_\odot$  \\
$M_{H_2}/M_\star$ &  $0.41^{+0.04}_{-0.12}$   \\
\\
$M_{\rm dust}$ & {$<1.9\times10^{8}~M_\odot$}~(IR)\\
 & {$<2.0\times10^{8}~M_\odot$}~(1.3~mm continuum)\\

$M_{H_2}/M_{\rm dust}$    & {$\gtrsim400$} \\
\\
$\tau_{\rm dep}= M_{H_2}/{\rm SFR}$ & $(1.3\pm0.4)$~Gyr     \\
 \end{tabular}
\end{center}
 \caption{Stellar, gas, and dust properties of 3C~186 system.} 
\label{tab:gas_properties}
\end{table}

\section{Discussion and Conclusions}\label{sec:discussion_conclusions}
The scenario that emerges from our analysis is that we detect a large molecular gas reservoir over a $\sim16$~kpc scale region, {likely associated with a {disk-like structure} of a massive galaxy. Similar complex morphology and kinematics are observed both in distant and local cluster core galaxies \citep[e.g.,][]{Russell2019,Noble2019}. With this work} we confirm both the spatial and velocity offset, found in previous studies, with respect to the {BLR of the} QSO. These findings confirm the 3C~186 system as an excellent candidate for a kicked super-massive black hole, {resulting} from a strong gravitational wave recoil as two SMBHs coalesced after the merger of their host galaxies.
The present study thus reports the first confirmation of a GW recoil candidate via high-resolution mm observations.

Rest frame optical-UV SED modeling by \citet{Morishita2022} found that the SF blob dominates the total star formation of the system, with an estimated ${\rm SFR}=(65\pm20)~M_\odot/$yr, well in agreement with the upper limit to the total ${\rm SFR}<80~M_\odot$/yr, as estimated by \citet{Podigachoski2015} by modeling the {near- to far-IR photometry} of 3C~186. Conversely, the SED analysis by \citet{Morishita2022} shows that the SF blob contributes to only 4\% to the total stellar mass  $M_\star=(2.04^{+0.59}_{-0.18})\times10^{11}~M_\odot$ of the system. 

The estimated SFR is consistent with that of MS field galaxies SFR$_{\rm MS} = 76~M_\odot$/yr \citep{Speagle2014} with $M_\star$ {and redshift equal to those of 3C~186.}  These findings support our choice for a Galactic CO-to-$H_2$ conversion factor, while a lower $\alpha_{\rm CO}$ would be more appropriate for more star forming galaxies with SFR$>3~$SFR$_{\rm MS}$. The system is not only massive, but also gas rich. Our observations yield a $H_2$-to-stellar-mass ratio of $M_{H_2}/M_\star=0.41^{+0.04}_{-0.12}$, well in agreement with the MS values for field galaxies $(M_{H_2}/M_\star)_{\rm MS}=0.30\pm0.20$, with $M_\star$ {and redshift equal to those of 3C~186 \citep{Tacconi2018}.}

{Under the assumption that the $H_2$-to-stellar mass ratio reflects that of field galaxies we can express the ratio as $M_{{\rm H}_2}/M_\star\sim0.1\,(1+z)^2$ \citep{Carilli_Walter2013}. 
This yields $M_\star<1.6\times10^{10}~M_\odot$, which is the upper limit to the stellar mass of the QSO host galaxy, in the case where the quasar host is superimposed, in projection, to the brighter offset galaxy. The QSO host should then be of low-mass, which is quite unlikely for a radio-loud AGN at the cluster center because i) the hosts of radio loud QSOs are typically massive ellipticals with $M_\star$ largely exceeding $10^{10}~M_\odot$ and, similarly, ii) cluster cores are predominantly populated by massive galaxies. The possibility that the 3C~186 QSO is hosted by an under-massive galaxy is thus quite unlikely, as previously discussed by \citet{Chiaberge2017} on the basis of the  $M_\star$ - $M_\bullet$ relation.}

{Furthermore,} we estimate a depletion time-scale of $\tau_{\rm dep}=M_{H_2}/{\rm SFR}=1/{\rm SFE}=(1.3\pm0.4)$~Gyr, slightly higher than that predicted for field MS galaxies, $\tau_{\rm dep,MS}=(0.93^{+0.18}_{-0.15})$~Gyr, at given $M_\star$ and redshift \citep{Tacconi2018}. {This implies}  a relatively low star formation efficiency (SFE) for the system: part of the $H_2$ reservoir may be recently accreted, and is not being converted effectively into stars: an interpretation that is in agreement with the occurrence of a past merger (Sect.~\ref{sec:Mdust}). 

{Throughout our analysis we adopted a Galactic CO-to-$H_2$ conversion factor. The possibility to have instead a lower $\alpha_{\rm CO}=0.8~M_\odot~({\rm K}~{\rm km}~{\rm s}^{-1}~{\rm pc}^2)^{-1}$, typical of starbursts, cannot be firmly excluded. However, it would lead to a relatively low $M_{H_2}/M_\star\simeq0.08$ and a short $\tau_{\rm dep}\simeq0.2$~Gyr, with respect to the MS predictions.}

While the bulk of the CO emission is cospatial with the host galaxy isophotal center, we observe an offset with respect to the {North-East} star-forming blob, as often found in high-$z$  brightest cluster galaxies \citep[BCGs,][]{Strazzullo2018,DAmato2020}. 
It is likely that a large fraction of the rest-frame UV emission of the galaxy is obscured by dust, and thus not seen in Fig.~\ref{fig:HST_image}, {consistent} with the non-negligible {far-IR} emission reported by \citet{Podigachoski2015}. 

It is also possible that we are tracing only the most dense and central regions of the gas reservoir, while lower-J CO emission could reveal a more extended gas reservoir. We find indeed evidence for diffuse S/N$\lesssim3$ emission that is coherently distributed along the periphery of the emitting region (Figs.~\ref{fig:maps}a, \ref{fig:HST_image}). 
An alternative {explanation} is that part of the observed extended emission is associated {with diffuse} circum-galactic medium (CGM). However, we think it is less likely, as we would expect much larger line widths, more complex line kinematics, and extended emission over tens of kpc or more \citep{Emonts2016,Ginolfi2017,Cicone2021}.

Overall, our study reveals a large molecular gas reservoir, likely associated with a rotating disk, but still in formation, as suggested by the observed complex kinematics and multiple components. {Indeed, the highest velocity gradients (Fig.~\ref{fig:maps}b) are found between the central part of the emitting region and the adjacent northen and southern components. The latter recedes along the line of sight with respect to both the bulk of the gas reservoir and the central quasar, while the end of the {southern component} appears blueshifted. These gradients suggest that the 
{latter} may be a disk component that is torqued (e.g., a tidal tail), possibly due to past dynamical interactions. Interestingly, similar complex features and kinematics are often observed in merger remnants \citep{Ueda2014}. It is thus possible that the observed molecular emitting gas is associated with a disk in formation.} 

{The two morphological extensions of the main CO emitting region, i.e. the {\it N} and {\it S} components} (Sect.~\ref{sec:results}), are thus reminiscent of a strong environmental processing of the gas occurring in the cluster core. The cool-core cluster environment may favor the condensation and the inflow of the intra-cluster medium (ICM) gas towards the cluster center, as found only for some local \citep{Salome2006, Tremblay2016} and intermediate-$z$ BCGs \citep{Castignani2020c}.
Such a condensation {is manifested} by means of the significant molecular gas reservoir that we observe in the 3C~186 system, {possibly regulated by radio-mode AGN feedback originated by the non-thermal emission of the QSO.} {In 3C~186} we thus witness the assembly of a high-$z$ progenitor of local BCGs in great detail, at much higher resolution than recent studies \citep{Castignani2020a,DAmato2020}.

{In conclusion, the projected and spectral  offsets of the molecular gas reservoir with respect to the BLR of the QSO, as well as the complex morphology and kinematics of the 3C~186 system, presented in this study, strongly support the scenario that 3C~186 is a result of a GW recoil following a past merger. The system is an excellent target for {follow-up observations with the} {\it James Webb Space Telescope}.}  



\begin{acknowledgements}
{We thank the anonymous referee for helpful comments which contributed to improve the paper.}
This work is based on observations carried out under project number W20CO with the IRAM NOEMA Interferometer. IRAM is supported by INSU/CNRS (France), MPG (Germany) and IGN (Spain).
GC thanks Melanie Krips and all IRAM staff for the observations and help concerning the data reduction.
{GC acknowledges the support from the grant ASI n.2018-23-HH.0.}
The authors thank Bryan Hilbert and Erini Lambrides for helpful discussion.
\end{acknowledgements}








\onecolumn
\begin{appendix}
\section{{NOEMA maps and spectra}}\label{app:COspectra}
{Here we provide NOEMA maps and spectra. We refer to the Sects~\ref{sec:results} and \ref{sec:discussion_conclusions} for details and discussion.}

\begin{figure*}[h!]
\begin{center}
 \subfloat[]{\includegraphics[width=0.41\textwidth]{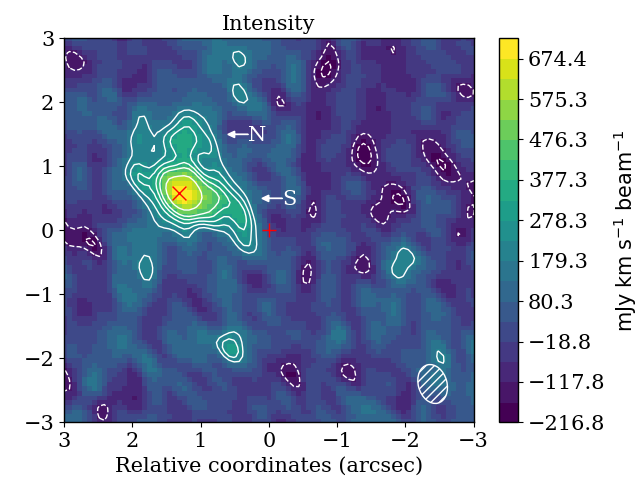}}\\
 \subfloat[]{\includegraphics[width=0.41\textwidth]{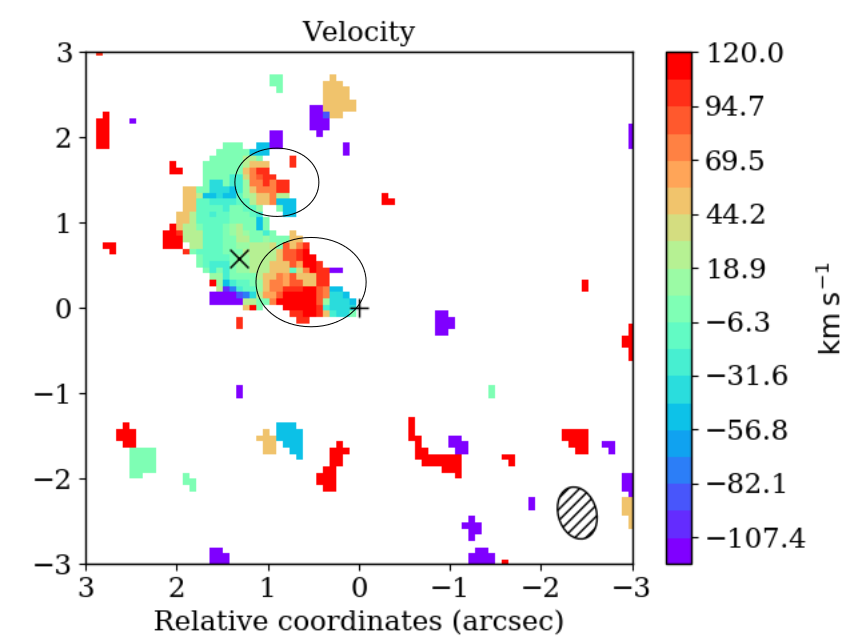}}
\vspace{-0.2cm}\subfloat[]{\includegraphics[width=0.41\textwidth]{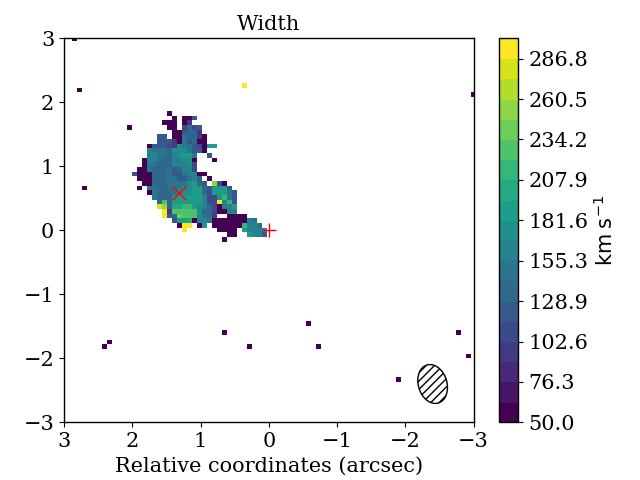}}\\
 \vspace{-0.2cm}\subfloat[]{\includegraphics[width=0.41\textwidth]{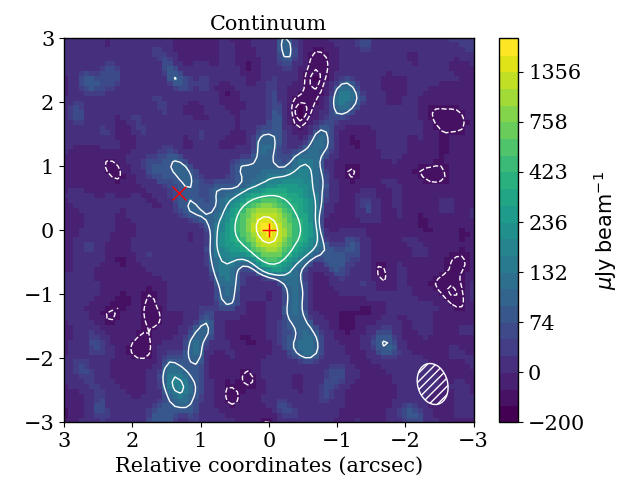}}
 \subfloat[]{\includegraphics[width=0.41\textwidth]{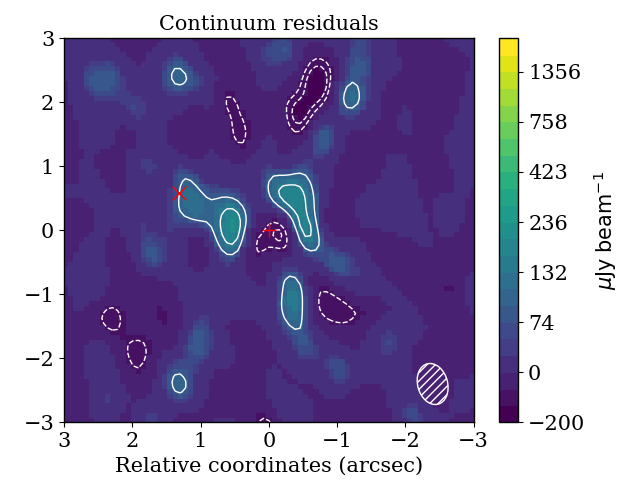}}
 \end{center}\vspace{-0.4cm}
 \caption{NOEMA maps. (a) Intensity, (b) velocity, and (c) velocity dispersion CO(4$\rightarrow$3) maps derived from the moments 0, 1, and 2, respectively. Panel~(d) shows the continuum map in the lower sideband (LSB), {while panel (e) displays the map of the residuals, once the central point-like component has been removed.}
Coordinates are reported as angular separations from {the phase center, i.e.,} the 3C~186 QSO coordinates. {The phase center and the CO intensity peak are marked with the symbols + and x, respectively.} The northern (N) and the southern (S) {components} are highlighted with arrows (panel~a) and ellipses (panel~b). In the panels~(a-c) the velocity range considered corresponds to the velocity support [-350; 330]~km~s$^{-1}$ of the CO(4$\rightarrow$3) line, while in panels (d, {e}) all 7.7~GHz bandwidth of the LSB is used.  Contour levels are superimposed to both the intensity and continuum maps.  They correspond to significance levels of {-3, -2, 3, 4, 5, 6, 7, and 9$\sigma$} (panel~a) as well as {-3, -2, 3, 5, 13, and 45$\sigma$ (panels~d, e).}  {Dashed and solid contours refer to negative and positive significance levels, respectively.}  
 The dashed ellipses at the bottom right of each panel show the beam size.}
\label{fig:maps}
\end{figure*}

\begin{figure*}[h!]
\begin{center}
\subfloat[]{\includegraphics[width=0.47\textwidth]{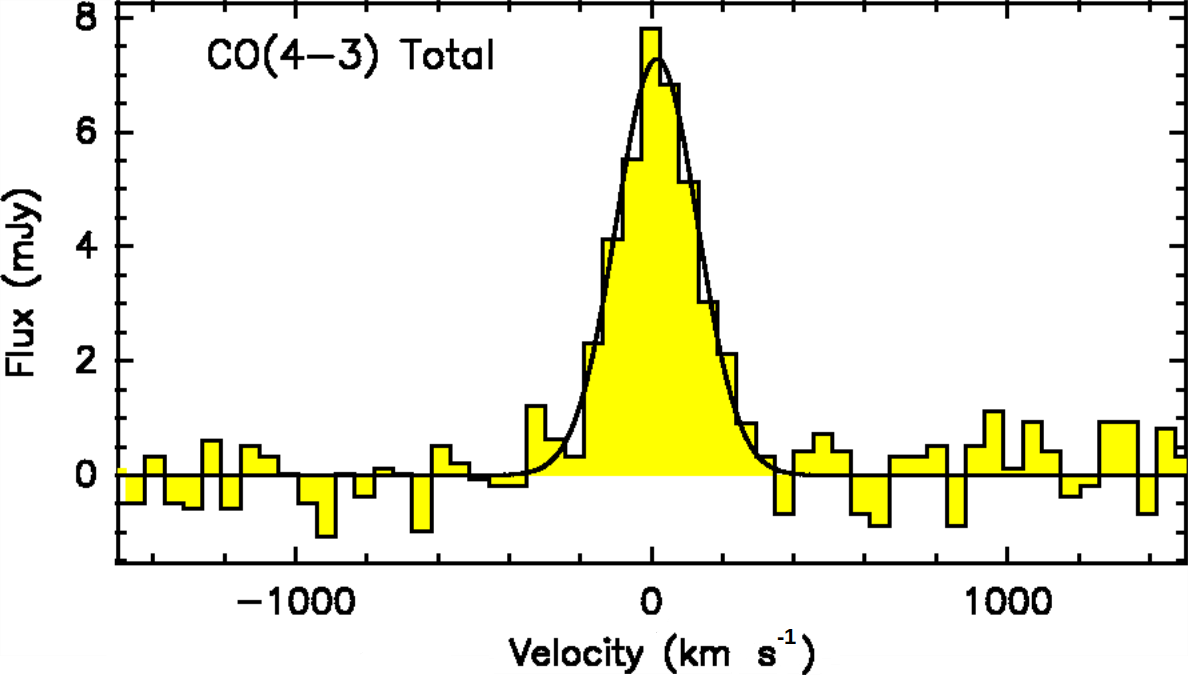}}
 \hspace{0.4cm}\subfloat[]{\includegraphics[width=0.47\textwidth]{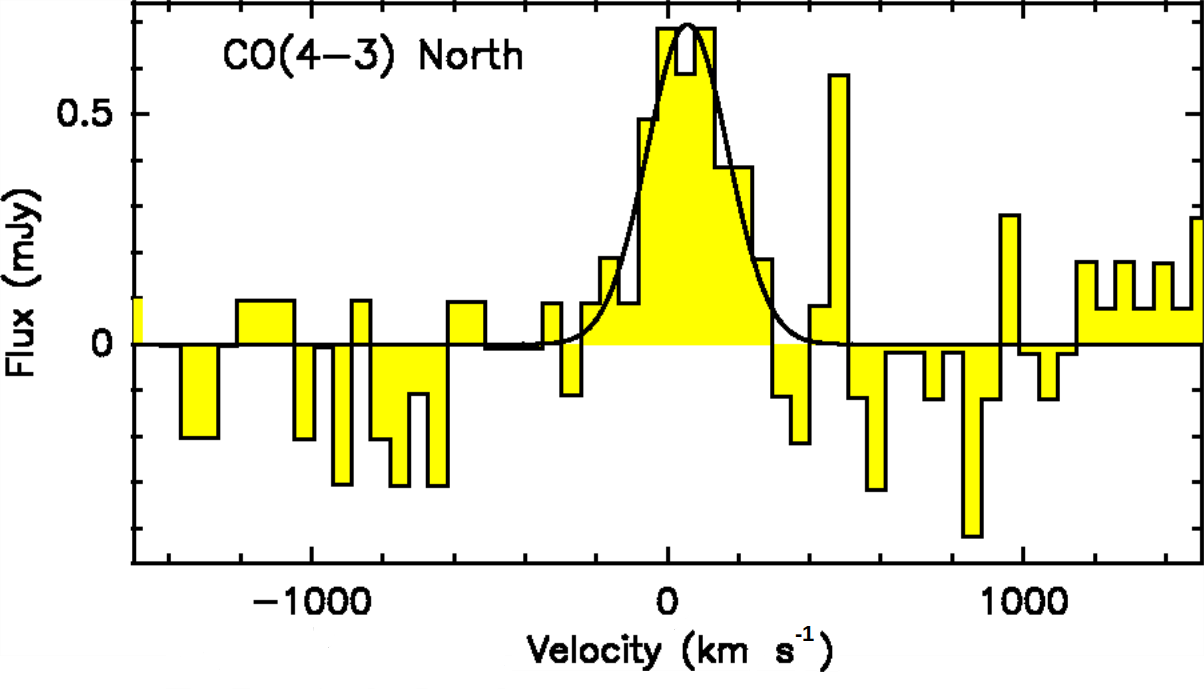}}\\ 
   \hspace{0.cm}\subfloat[]{\includegraphics[width=0.47\textwidth]{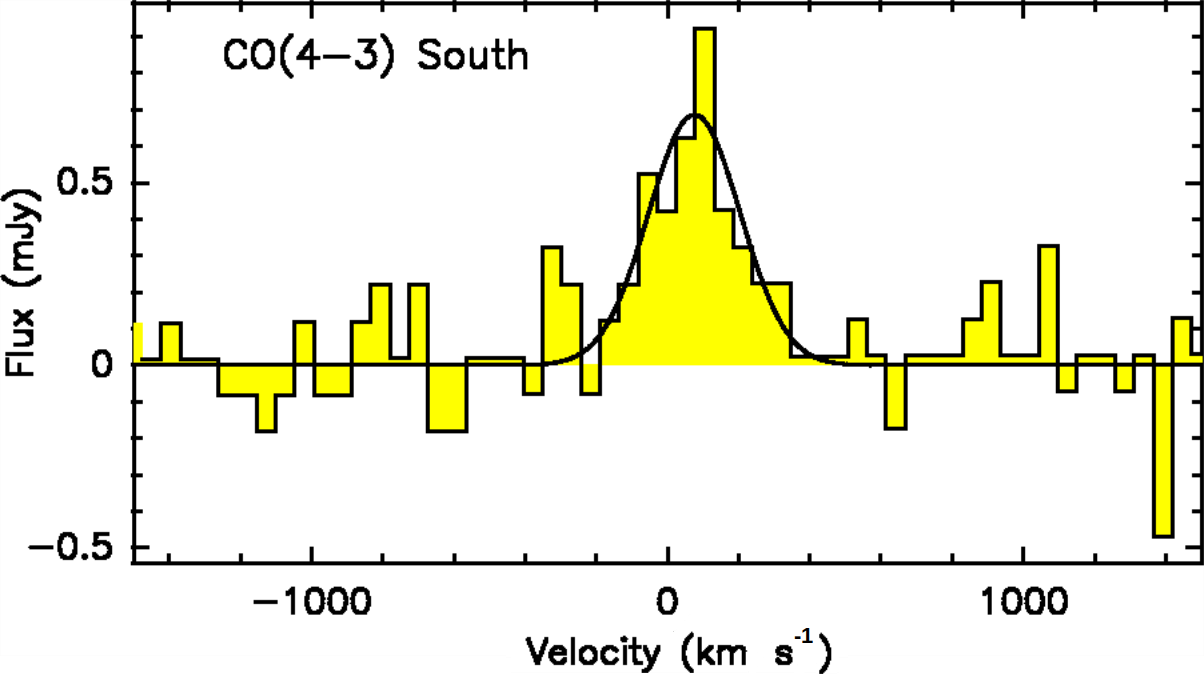}} 
   \hspace{0.4cm}\subfloat[]{\includegraphics[width=0.47\textwidth]{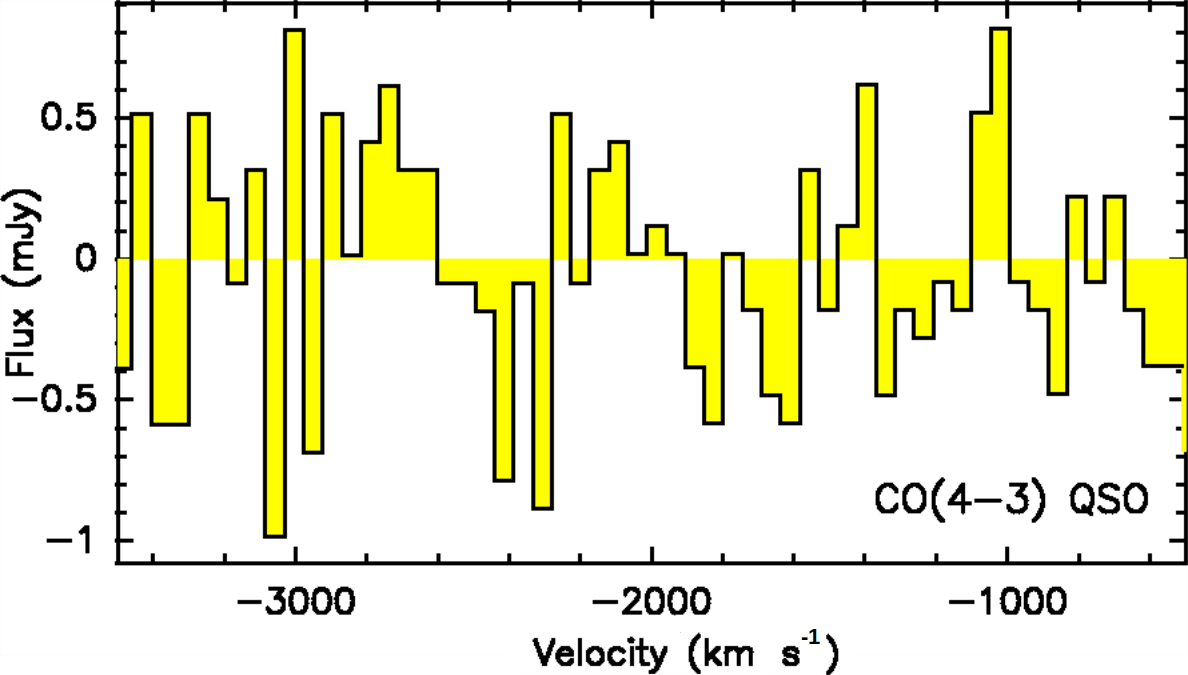}} 
\end{center}
 \caption{Baseline subtracted CO(4$\rightarrow$3) spectra for the total  emitting region (a), the northern component (b), the southern {component} (c), and the QSO (d). {For the {\it N} and {\it S} components we choose as regions of integration those denoted by the ellipses in Fig.~\ref{fig:maps}b, while for the QSO we chose a $1^{\prime\prime}\times1^{\prime\prime}$ region centered around it that fairly corresponds to the extent of the continuum emission (Fig.~\ref{fig:maps}d).}  In the panels a, b, and c the solid line shows the Gaussian best fit to the CO(4$\rightarrow$3) line. Velocities, shown in the x-axis, are evaluated with respect to the NLR redshift. The BLR is blueshifted to a velocity of -2035.7~km~s$^{-1}$.}
\label{fig:CO43spectrum}
\end{figure*}
\end{appendix}
\end{document}